# Precise influence evaluation in complex networks

**Bingyu Zhu[1], Qingyun Sun[2], Jianxin Li[2], Daqing Li[1]***

1 School of Reliability and Systems Engineering, Beihang University, Beijing 100191, China

2 School of Computer Science and Engineering, Beihang University, Beijing 100191, China

Email: daqingl@buaa.edu.cn

## Abstract

Evaluating node influence is fundamental for identifying key nodes in complex networks. Existing methods typically rely on generic indicators to rank node influence across diverse networks, thereby ignoring the individualized features of each network itself. Actually, node influence stems not only from general features but the multi-scale individualized information encompassing specific network structure and task. Here we design an active learning architecture to predict node influence quantitively and precisely, which samples representative nodes based on graph entropy correlation matrix integrating multi-scale individualized information. This brings two intuitive advantages: (1) discovering potential high-influence but weak-connected nodes that are usually ignored in existing methods, (2) improving the influence maximization strategy by deducing influence interference. Significantly, our architecture demonstrates exceptional transfer learning capabilities across multiple types of networks, which can identify those key nodes with large disputation across different existing methods. Additionally, our approach, combined with a simple greedy algorithm, exhibits dominant performance in solving the influence maximization problem. This architecture holds great potential for applications in graph mining and prediction tasks.

## Introduction

The variation of the local features in the network might spread to part or even the whole network[1-7]. These emergent influences can result in explosive catastrophes [5-11], such as traffic gridlock[12,13], rumor spreading[14-16], power grid collapses[17-20], disease outbreaks[21,22]. For example, an overloaded transmission line triggered a cascading failure in the Turkish power grid, impacting approximately 75 million people and causing at least $700 million economic losses in 2015[23]. And in the biomedicine field of epilepsy, deep brain stimulation on the most influential downstream "propagation points" can prevent onward spread of seizure activity by desynchronizing the epileptogenic network, which has become effective and available treatment options to reduce seizure burden for selected patients with drug-resistant epilepsy[24]. On the other hand, people also harness this explosive influence for purposes of marketing and message dissemination[25-29]. For instance, at 2012, an Australian public campaign video, "Dumb Ways to Die" made by Metro Trains in Melbourne went viral on social media after it was released in November 2012 and had been generated at least $50 million worth of global media value within two weeks by estimate, contributing to a 20 percent reduction in "near-miss" accidents compared to the annual average[30-32]. Hence, for mitigating influence of cascading failures or optimizing influence of opinion propagation[33-36], the identification of key nodes with significant spreading influence in large-scale networks holds paramount importance[37-40]. However, it remains challenging to precisely identify these critical structures with strong propagative influence.

Identifying influential nodes contains two manifolds. Firstly, it involves precisely evaluating the influence of single nodes[39-46], and secondly, it entails discovering a group of nodes that can maximize their collective influence[47-50]. Currently, the evaluation methods for node influence include network structure analysis and dynamic simulation[51-53]. Network analysis methods typically rank the influence of nodes based on structure centrality of nodes using global measures (k-shell[37], H-index[45], Vote Rank[38]), local measures (degree[54], DCL[44]) or semi-local measures (CI[49], Local Rank[46], LID[42]). Then key nodes are directly selected as seed nodes for influence maximization strategies based on the ranking. However, these approaches may miss those nodes with large influence, yet embedded in elusive structure features. Although simulation methods can precisely evaluate the node influence, the computation time for identifying key nodes is immeasurable for large-scale networks due to the probabilistic characteristic of the propagation models. Recently, the application of deep learning has shown outstanding performance in the field of complex networks, such as FINDER[55] and GDM[56] and there have been algorithms designed for ranking network influence. One of them is the RCNN, which utilizes convolutional neural networks on local neighborhoods to predict nodes spreading ability[57]. Another method called Graph-based LSTM uses the Long Short-Term Memory model to process the entire graph and considers the proximity of node features[58]. However, because of their non-adjustable training, these deep learning methods still cannot achieve precise prediction of influence in different networks.

Therefore, the above methods of qualitative ranking exhibit unstable performance in different networks, uncovering the challenge of achieving precise evaluation across diverse networks. Precise influence evaluation across different networks requires quantitative precise estimation of node influence values. However, the transferability and generality of methods in diverse networks necessitate the consideration of network individualized features, due to the sensitivity of influence to the variation of structure. Graph learning methods incorporating active learning process may address this issue[59-61], as active learning allows for further adjustment to the prediction model based on network individualized features, adapting influence prediction for new networks. This brings two intuitive advantages: (1) discovering potential high-influence but weak-connected nodes which are usually ignored in existing methods, (2) improving the influence maximization strategy by deducing influence interference quantified by precise influence values.

In this study, we introduce a novel **A**ctive **L**earning framework based on **G**raph **E**ntropy (**ALGE**), enabling precise influence evaluation in various types of networks, as depicted in Fig. 1. In detail, a graph neural network is employed as the underlying predictor, which effectively establishes the mapping relationship between node features and influence. And then, the individualized structural features of the target network (a representative subset of nodes) are sampled by active learning strategy based on graph entropy (for more details, please refer to Methodology). Ultimately, the simulated influence of the representative nodes fine-tunes the graph neural network predictor to achieve precise influence evaluation. Although the active learning process requires minimal simulations, this low-cost learning approaches yield excellent generalization capabilities across various types of networks， such as social networks, biological networks and infrastructure networks, enabling quantitative prediction of node influence in experiments. As a result, a simple and low-complexity greedy algorithm based on the node influence evaluation is developed to tackle the influence maximization problem. This algorithm can intuitively avoid the overlap propagation coverage of multiple key nodes and thus maximize the collective influence. Experimental results demonstrate that this algorithm outperforms most methods on diverse networks.

## Result

For a network, each node has its own influence ability in the stochastic spreading process modeled by the standard SIR models, which can be quantified by the whole infected nodes when a particular node is the sole initial infected node and all other nodes are susceptible[62] , as shown in Fig. 1a. Precise influence evaluation across different networks requires not only ranking prediction of node influence but also the quantitative estimation of node influence values, which is the number of nodes infected by the initial one. Here, we achieve precise influence evaluation in various types of networks, by a novel active learning framework based on graph entropy (ALGE). The precise influence evaluation framework proposed consists of three parts: (1) pre-training based on task information and artificially synthetic networks, (2) active learning based on real network structural information, (3) prediction of influence on real networks, as shown in Fig. 1b.

In the pre-training part, the initialized GNN model is trained to a universal basic prediction model by the simulation data on the synthetic network. Due to the small scale of the synthetic network, it allows for low-cost simulation to obtain labeled data for the true influence. And the predictor is a graph deep learning model, comprised of a graph convolutional module and a regression module. The graph convolutional module applies 3 graph attention layers with 8 heads, which can capture the features of nodes and their neighborhoods. This structure has been proven to better capture attention weights [63], which means the mutual impact between influence of nodes. Indeed, the pre-trained predictor, named Basic Model (ALGE-B) has demonstrated good performance in ranking node influence on real networks (seen in Fig. 2a). After learning the general structural features by the basic pre-trained predictor on synthetic network data, we utilize the active learning (AL) to further learn the individualized global features in the second part, aiming to achieve precise prediction of influence values in real networks. During the AL process, we first construct an entropy-correlation network based on the relative entropy similarity[64,65] among nodes in the graph where each edge symbolizes the similarity between nodes. Subsequently, we select the node similar to the most nodes in the entropy-correlation network, add it to the representative node set, and remove the selected node and its adjacent nodes. By iteratively repeating the selection-removal process until no nodes remain in the network, the representative node set is obtained. By simulation of the representative node set of the entropy-correlation network, labeled data for representative nodes of the real network is obtained. Finally, after fine-turning the basic predictor using the real influence values of the labeled representative nodes, we obtained a comprehensive prediction model (ALGE-C), allowing for precise and quantitative evaluation of node influence in the real network (seen in Fig. 2a, 2b). More details about our method can be seen in Methodology.

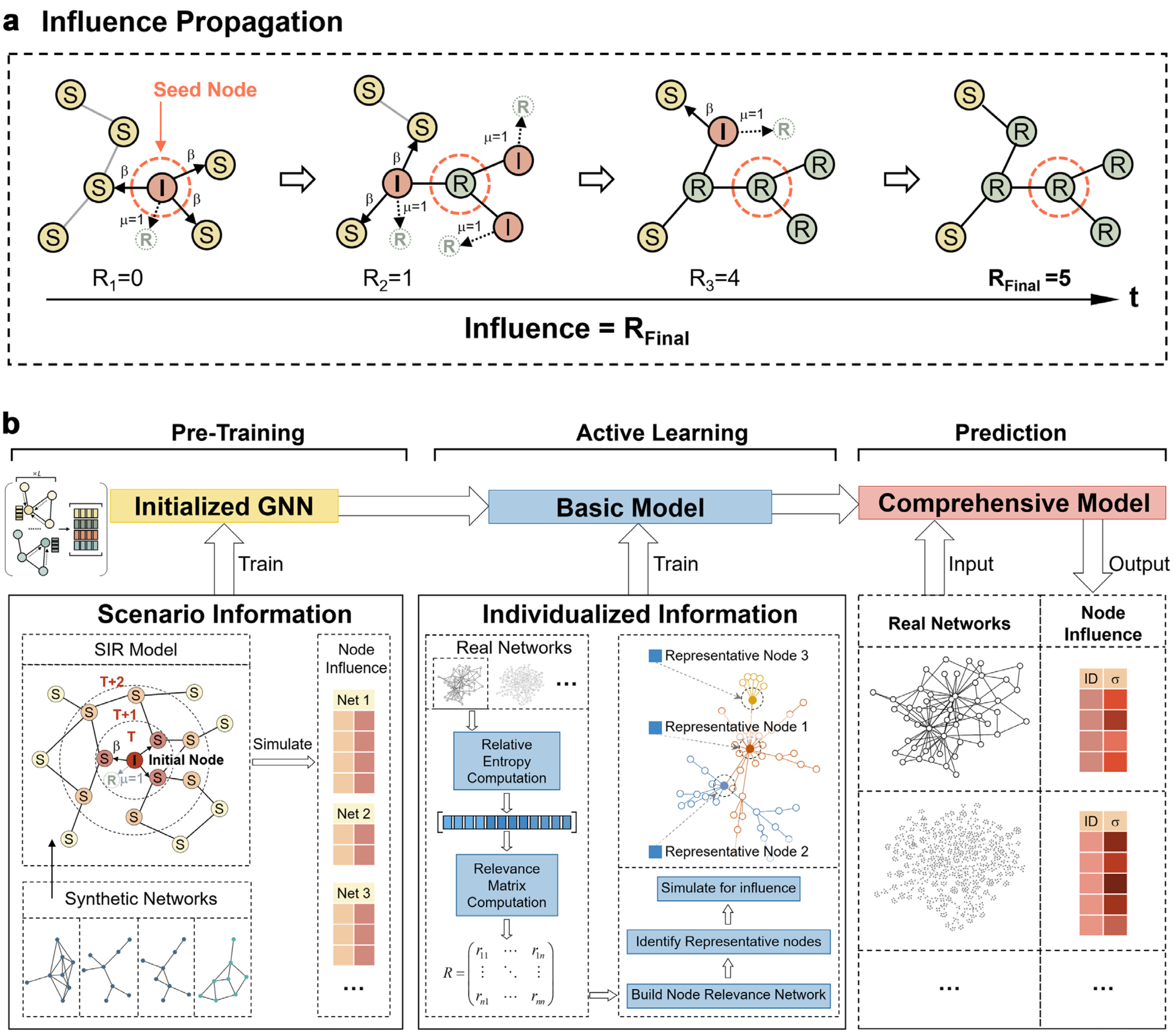

**Fig. 1 Node Influence Prediction. a** The definition of node influence. The diffusion model is the standard SIR model usually employed to analyze the node influence, where S, I, R respectively denote susceptible, infected, and recovered nodes, and $\beta$ and $\mu$ represent the probability of infection and recovery respectively[22]. For simplicity, $\mu$ is commonly set to 1[45,47]. In the SIR model, the node influence is defined as the number of final recovered nodes, when a particular node is the sole initial infected node and all other nodes are susceptible[47]. And the true node influence is calculated by the average of the final number of infected nodes from 1,000 simulations under the seed node[45]. **b** The Precise influence evaluation framework with Active Learning based on Graph Entropy (ALGE). The framework comprises three parts: (1) pre-training based on task information and artificially synthetic networks, (2) active-transfer learning based on real network structural information, (3) prediction of influence on real networks. The framework can find influential nodes based on the multi-scale information encompassing structure and task.

## Precise influence evaluation in diverse networks

Precise influence evaluation helps to identify potential key nodes, and provide solutions for decision-making problems in networks. And the precise influence evaluation contains two parts: ranking the node based on the influence precisely, and estimating the influence values precisely. Significantly, our method is competent for these two parts and the performance of precise ranking of node influence is illustrated first below. Our proposed method demonstrates higher level of precision in ranking node influence, comparing with existing approaches in predicting node influence, as shown in Fig. 2a and Table 1. In Fig. 2a, we compare different methods in ranking node influence with correlation between true rank and predicted rank in the malaria gene network as an example. It is evident that the ALGE-C method exhibits the best fit in node ranking. Furthermore, to verify the generality of ALGE applied in influence ranking, we conduct experiments on 26 different real networks to rank the influence of nodes (as seen in Table 1). Kendall's coefficients (refer to Methodology for details) between prediction results and the true rank are calculated to evaluate the performance of ALGE compared with the existing 8 algorithms. A higher Kendall's coefficient indicates better performance in ranking prediction closer to the true rank. It can be found that our comprehensive model (ALGE-C) outperforms other algorithms in Table 1. Furthermore, the ALGE-C model gains the capability to precisely estimate the node influence values quantitatively (shown in Fig. 2b), which is crucial as it provides valuable information for various decision-making processes. In our study, we attempt to predict the influence values in 34 different real networks, and compare the accuracy of the predictions before (ALGE-B) and after retraining (ALGE-C). As shown in Fig. 2c, the ALGE-C model retrained based on active learning significantly improves the accuracy of predicting node influence values. By contrast, the fundamental model (ALGE-B) can only provide a general trend in predicting node influence values, with significant deviations in numerical accuracy (see Supplementary Figure 2 for more results).

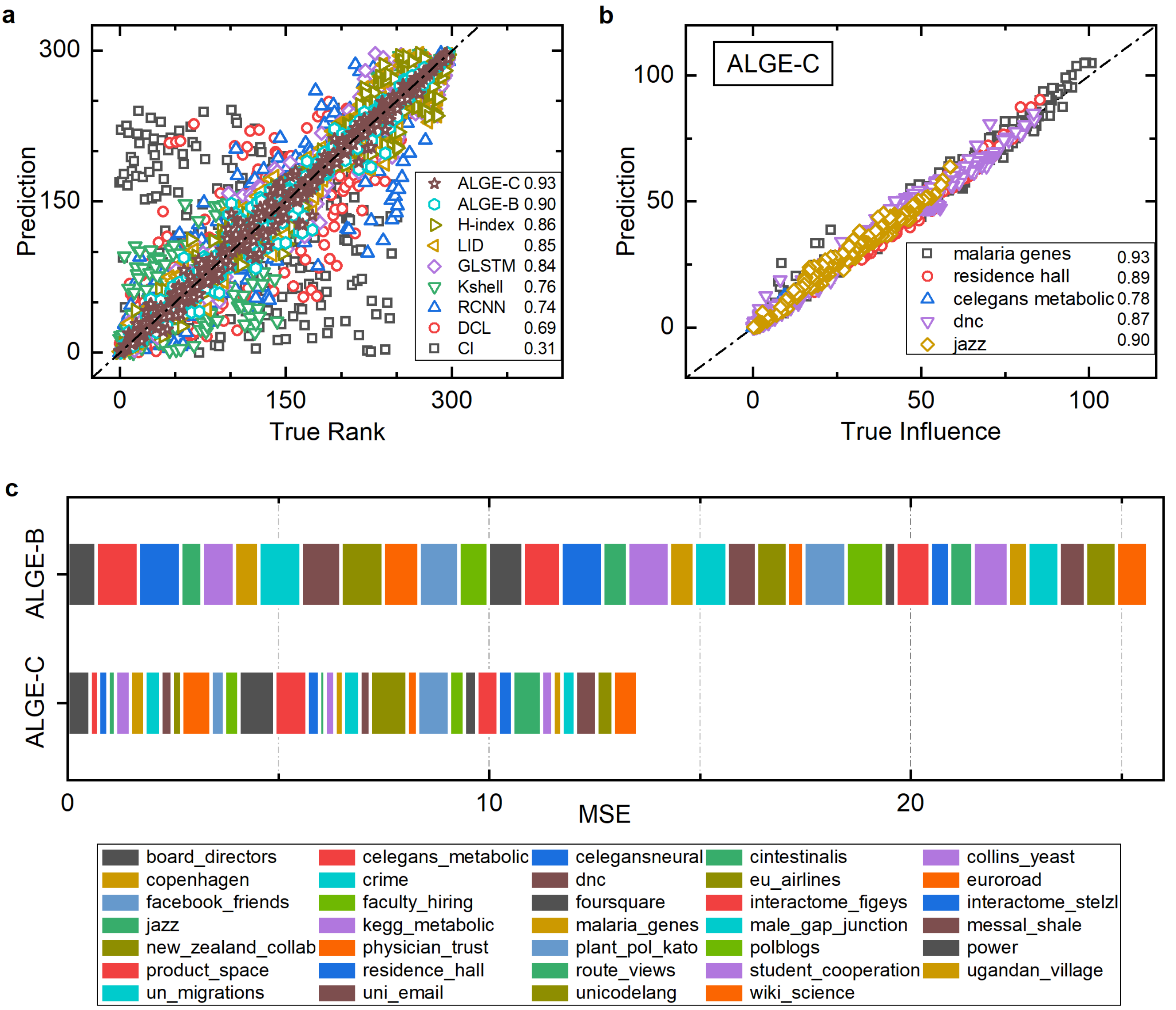

**Fig. 2 The performance of our prediction framework. a** The comparison between different methods to predict the rank of node influence in the malaria genes network. The horizontal axis represents the true rank obtained by 1,000 independent simulations, while the vertical axis represents the predicted ranking. ALGE-C exhibits the best fit in node ranking. **b** The influence values of ALGE-C compared with simulation. Influence values estimated by ALGE-C method fits simulation result well. **c** The comparison between ALGE-C and ALGE-B in influence value prediction. The horizontal axis represents the mean squared error (MSE) of the predicted results. ALGE-C improves the accuracy of predicting node influence values from ALGE-B. And the basic features and reference of these real networks used in our analyze can be seen in Supplementary Table 1.

**Table 1 Kendall's coefficients between influence rank of SIR simulation and algorithms. Our comprehensive model (ALGE-C) outperforms other algorithms in influence ranking.**

| Networks | CI | k-shell | H-index | LID | DCL | NCVR | RCNN | GLSTM | ALGE-C |
|---|---|---|---|---|---|---|---|---|---|
| Board directors | 0.3927 | 0.3346 | 0.3301 | 0.3119 | 0.3501 | 0.2304 | 0.3979 | 0.3101 | **0.6777** |
| Celegans metabolic | 0.1747 | 0.6577 | 0.6464 | 0.6308 | 0.3122 | 0.1362 | 0.5101 | 0.4768 | **0.7843** |
| Celegansneural | 0.2349 | 0.6976 | 0.7667 | 0.7306 | 0.3460 | 0.3211 | 0.6564 | 0.7184 | **0.7873** |
| Cintestinalis | -0.358 | 0.8371 | 0.8982 | 0.8784 | 0.6785 | 0.1878 | 0.7742 | 0.8656 | **0.9088** |
| Collins yeast | 0.6633 | 0.6472 | 0.6622 | 0.6517 | 0.6585 | 0.1399 | 0.6685 | 0.6209 | **0.8322** |
| Copenhagen | 0.6355 | 0.8014 | 0.8242 | 0.7849 | 0.5583 | 0.1772 | 0.6988 | 0.7480 | **0.8995** |
| Crime | 0.3819 | 0.2648 | 0.2909 | 0.2865 | 0.4721 | 0.1691 | 0.4642 | 0.2410 | **0.7294** |
| Dnc | 0.3816 | 0.5745 | 0.5775 | 0.5740 | 0.0853 | 0.3031 | 0.6016 | 0.5371 | **0.7766** |
| Eu airlines | 0.3796 | 0.8299 | 0.8141 | 0.8045 | 0.2972 | 0.5237 | 0.6000 | 0.7881 | **0.8667** |
| Euroroad | 0.6636 | 0.4450 | 0.4391 | 0.4373 | 0.4855 | 0.1348 | 0.4444 | 0.4317 | **0.7024** |
| Facebook friends | 0.6852 | 0.6620 | 0.6903 | 0.6932 | 0.5547 | 0.1862 | 0.6098 | 0.6731 | **0.8728** |
| Foursquare | 0.6414 | 0.6064 | 0.5462 | 0.5381 | 0.7203 | 0.3380 | 0.5386 | 0.5506 | **0.8396** |
| Interactome stelzl | 0.5863 | 0.5824 | 0.5399 | 0.5284 | 0.7562 | 0.3444 | 0.5478 | 0.4881 | **0.8489** |
| Jazz | -0.003 | 0.7794 | 0.8586 | 0.7534 | 0.6770 | 0.3636 | 0.6656 | 0.8141 | **0.9027** |
| Kegg metabolic | 0.4191 | 0.6430 | 0.6549 | 0.6520 | 0.3843 | 0.2924 | 0.4122 | 0.5389 | **0.7112** |
| Malaria genes | 0.3117 | 0.7673 | 0.8697 | 0.8587 | 0.6954 | 0.0126 | 0.7417 | 0.8401 | **0.9331** |
| Male gap junction | 0.7701 | 0.6927 | 0.6789 | 0.6655 | 0.6265 | -0.113 | 0.6946 | 0.6383 | **0.8792** |
| Messal shale | 0.4539 | 0.7218 | 0.7585 | 0.7397 | 0.2972 | 0.1884 | 0.6561 | 0.6686 | **0.8691** |
| Physician trust | 0.6004 | 0.6678 | 0.7984 | 0.8105 | 0.4633 | 0.3958 | 0.6074 | 0.7831 | **0.8497** |
| Power | 0.4581 | 0.2874 | 0.3102 | 0.3079 | 0.3174 | 0.0960 | 0.4440 | 0.3073 | **0.6047** |
| Product space | 0.6748 | 0.5937 | 0.5945 | 0.5870 | 0.6574 | 0.1893 | 0.6757 | 0.5189 | **0.7467** |
| Residence hall | 0.0621 | 0.6902 | 0.8308 | 0.7920 | 0.5589 | 0.1406 | 0.6740 | 0.8144 | **0.8885** |
| Student cooperation | 0.7540 | 0.2298 | 0.3974 | 0.3978 | 0.4950 | -0.100 | 0.3167 | 0.3447 | **0.8359** |
| Ugandan village | 0.7671 | 0.6556 | 0.7736 | 0.7644 | 0.6049 | 0.2595 | 0.6503 | 0.7462 | **0.8588** |
| Uni email | 0.8151 | 0.8273 | 0.8318 | 0.8112 | 0.5888 | 0.4037 | 0.6399 | 0.7908 | **0.8489** |
| Wiki science | 0.7082 | 0.7880 | 0.7968 | 0.7703 | 0.5977 | 0.0912 | 0.6847 | 0.7404 | **0.9135** |

## Why ALGE could evaluate the node influence precisely

The network analysis framework of ALGE provides not only a more precise ranking of node influence but also a precise and quantitative description of the node influence values. To understand the advantage of ALGE model deeply, a new metric of the disputation is introduced to explain our method intuitively. The disputation of one node is defined as the average absolute difference between its rank index of given algorithms and its true rank index as shown in Fig. 3a (refer to Methodology for details). This shows that different algorithms give different rankings for influence evaluation. Then, we take the Power network as the case and calculate influence rank of the two nodes with 8 representative existing algorithms. The result in Fig. 3a shows two distinct nodes with the highest and lowest disputation indicating that these algorithms could yield extremely inconsistent rankings for node 4502 (DCL rank 50 and k-shell rank 3063). That is to say, due to the elusive structure information the node may exhibit different roles in different algorithm perspectives.

To explore the disputation effect for the rank, we calculate the disputation of 8 existing methods and divide the nodes into 10 groups based on their true influence rankings. The frequency distribution of disputation for each group is represented in Fig. 3b. In all 10 levels of influence rank, there exist significantly high disputations pervasively, and the nodes with exceptionally high disputations are concentrated in high-ranking and low-ranking positions. This implies that in the Power network, traditional methods are prone to severe wrong rank for some high-influence and low-influence nodes. To verify this, we compare the influence rank of nodes with top and bottom 100 disputations between the true and predicted values by different algorithms (Fig. 3c, 3d). While most of algorithms can give good estimation by definition for low disputation nodes, our method can well outperform the others for high disputation nodes. These result reveals the challenges faced in influence analysis. On one hand, ranking experience provided by empirical methods based on classical networks may be applicable to part of new network structures. On the other hand, there are other parts with elusive information leading to distinct judgements of different algorithms.

Here, we focus on two groups of nodes with distinct disputation in the Power network and compare their topologies (Fig. 3e). For the high-influence group, one high-disputation node 2583 is a three-degree node with a relatively sparse local neighborhood (12 second-order neighbors), prone to being misjudged as low-influence nodes. However, as the connecting center of three high-influence nodes, node 2583 with fast-growing neighbors especially at eighth-order actually has higher influence. This observation highlights that the node influence is a characteristic associated with multi-scale network structure. On the other hand, one low-disputation node 2582 is a typical high-degree hub with compact local structure (35 second-order neighbors), which can be easily identified as high-influence nodes by traditional methods, and thus have lower disputation. For the low-influence group, one high-disputation node 4502 appears as the hub node located at the center of a cluster. However, this cluster is relatively small (65 eighth-order neighbors) and positioned at the periphery of the overall network, leading to its actual low influence. The possible mistake of low-influence identification suggests that evaluating node influence cannot solely concentrate on the node position within local structures but should also consider the global position with respect to the entire network. Node 4651 with low influence and low-disputation has a low degree and locates at the margin of a small and remote cluster, resulting in hard propagation of influence.

The topological features of low-disputation nodes mentioned above are common and well represented by high-degree nodes in dense structure and low-degree nodes in sparse structures (long path or periphery)[37,54,66]. We refer to them as network common features, and traditional methods can reasonably rank node influence based on these topologies. Meanwhile, the topological features of high-disputation nodes above are more complicated. These unusual structures are actually individualized features present in each of real networks, which should be given equal consideration in influence analysis. Significantly, the argument above supports the advantage that our method can precisely evaluate the node influence. Through sampling the overall network structure by the active learning, our model (ALGE-C) can learn the effect of individualized network structures on local influence. Therefore, it enhances the accuracy of influence prediction on nodes with high disputation as shown in Fig. 3c, where our method ALGE-C achieves the highest Kendall's coefficient.

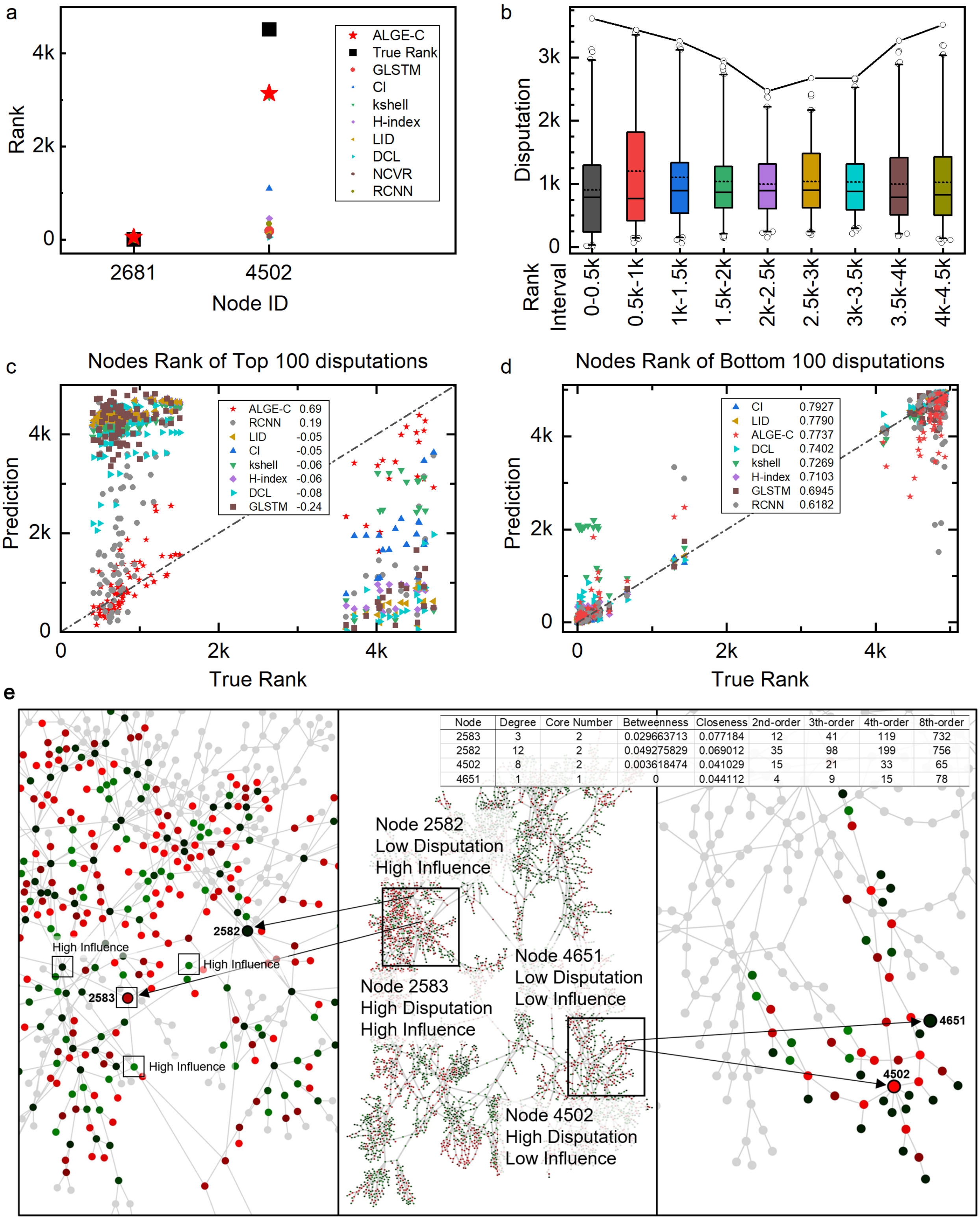

**Fig. 3 Advantage of ALGE shown in the Power network. a** The definition of disputation. The disputation is the dispersivity between true rank and predicted rank by given methods. The disputation of Node 2681 is much less than that of Node 4502. **b** The frequency distribution of disputation in different rank intervals. Each box demonstrates the node distributions in each rank interval, where dotted lines denote the mean value, solid lines indicate the median, and hollow points represent outliers. The nodes with higher and lower influence ranking have higher disputation. **c** The true and predicted influence rank of nodes with top 100 disputations. The result of our method ALGE-C aligns most closely with y=x (the black dot-and-dash line). The Kendall's coefficients of the 100 nodes are shown beside the method names. d The true and predicted influence rank of nodes with top 100 disputations. **e** Two groups of nodes with distinct disputation. The positions of nodes in the Power network are shown in the center. Nodes with disputation higher or lower than the mean value are colored by red and green respectively, where brighter nodes in the same color have higher disputation. On the left and right sides, the local topologies of the high-influence and low-influence node groups are respectively presented.

## Influence maximization problem based on our results

The influence maximization problem (IMP) refers to: selecting a set of initial nodes, and maximizing the number of ultimately influenced nodes under a given propagation model. IMP was first proposed by Kempe and has been proven to be an NP-hard problem[47]. This class of problems is typically solved by heuristic algorithms or greedy algorithms based on structural metrics[50]. Some algorithms consider the dispersion of seed nodes in the network as one of the criteria for selecting the seed node set[38,43,67] . When seed nodes are concentrated in the network, their influence ranges may overlap to some extent, resulting in the waste of influence. The overlap in the propagation process is manifested as their capability of propagation may be limited by the recovered nodes infected by seed nodes before. Therefore, estimating the influence values precisely is need to reduce interference between seed nodes intuitively. With the influence value, we can quantitatively analyze the mutual overlap limiting the infection abilities of seeds.

With our framework of ALGE-C model, we can capture the approximate influence ranges of all nodes in the network using breadth-first search (BFS) based on the influence values predicted. This enables us to directly identify overlap regions of influence between nodes. In this way, when solving IMP, the node influence can be efficiently leveraged by selecting seed nodes with high influence and minimal conflict (overlap) in their influence ranges. Based on this idea, we employ a simple greedy algorithm for seeds selection, named ALGE-Greedy. During ALGE-Greedy process, we iteratively select the highest influence node and update influence values of all nodes until the termination condition happens (refer to Methodology for details). We conduct simulations on 20 different real networks, and on average, our approach outperforms the others in IMP as shown in Fig. 4a. Taking the Power network as an instance, we compare the final infection size with different IMP methods as shown in Fig. 4b and 4c. It is evident that ALGE-Greedy method exhibits advantages over other methods across different seed sizes. Especially, during the propagation process with 15 seeds, ALGE-Greedy gains a significant advantage on infection size as time progresses, although lagging behind at first two steps.

For intuitive perspective, we illustrate the propagation process of seeds for different methods on the Power network. (Fig. 5a and 5b). The topology distribution of seed nodes indicates that accounting for both high influence and overlap effect numerically, our algorithm finds a more dispersive set than the traditional method CI, which selects seeds based on the optimal percolation. And at the first time step, the infection size of CI is higher than ALGE-Greedy. This is because seeds of CI with higher individual influence have less overlap among their infected regions at initial spreading and can leverage their influence capabilities. However, as time progresses, due to the compact distribution of CI seed nodes, significant overlap among their infected regions appears, resulting in poor collective influence. Indeed, this confirms the fact that the non-interference among seeds of ALGE-Greedy exert their influence efficiently though not generating instantaneous explosive influence as CI. In detail, we analyzed the disputation and infection size of the seed nodes selected by the ALGE-Greedy and CI in one simulation (Fig. 5 c), where some high-disputation nodes identified by ALGE-Greedy make contribution to influence maximization.

For deep insight of overlap effect, we quantify the overlap impact of seed nodes (refer to the Methodology for details) for collective influence. As shown in Fig. 5d, it is evident that most seeds of CI have high individual influence yet low contributions (actual infected number) to collective influence, while most seeds of ALGE-Greedy have medium individual influence and better contributions. We find that the influence decrease between individual influence and the contribution within collective influence exhibits an almost linear relationship with overlap (Fig. 5e). Evidently, the seed nodes of CI have higher overlap than those of ALGE-Greedy, resulting the actual influence of CI seeds decreases more in collective effect. In contrast, seeds of ALGE-Greedy with medium individual influence and lower overlap can better exert their individual influence, leading to higher collective influence. This confirms the viewpoint that in the IMP, not only the magnitude of seeds influence is important but also the dispersion of seeds is crucial. Our ALGE-Greedy for IMP provides new insights for the influence maximization problem.

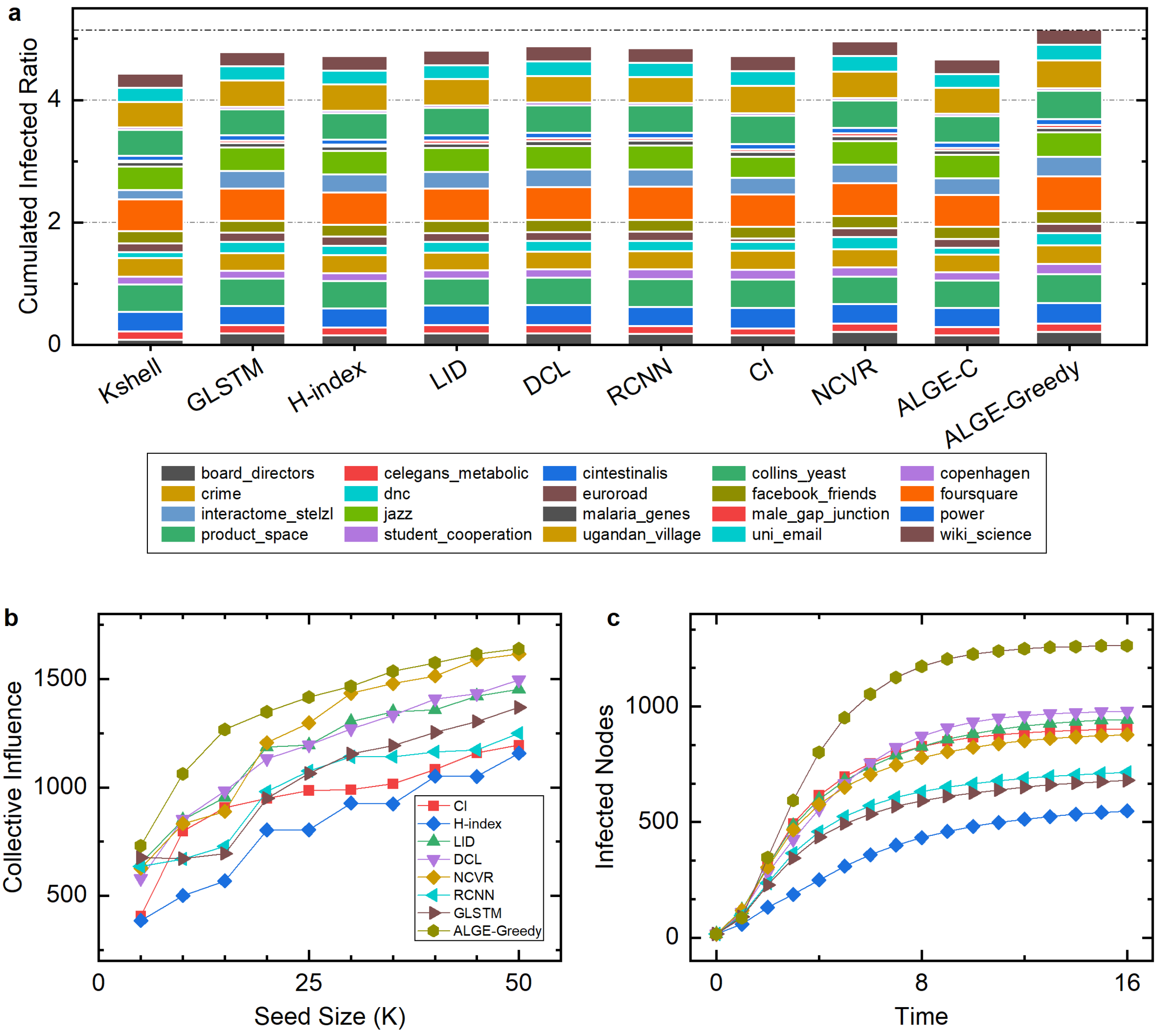

**Fig. 4 The universal performance of ALGE-Greedy in Influence Maximization Problem.** **a** The final infected ratio of different methods on 20 networks when the seed size is 15. The higher the better. More sizes of seeds can be seen in Supplementary Figure 5. **b** The collective influence of different methods on the influence maximization problem varies with different number of seed nodes in the Power network (see Supplementary Figure 6 for more networks). Each value of the collective influence is the average obtained from 1,000 independent simulations. **c** The increase of the infection size over time for different methods in the Power network (K=15). Each value of the infection size is the average obtained from 1,000 independent simulations.

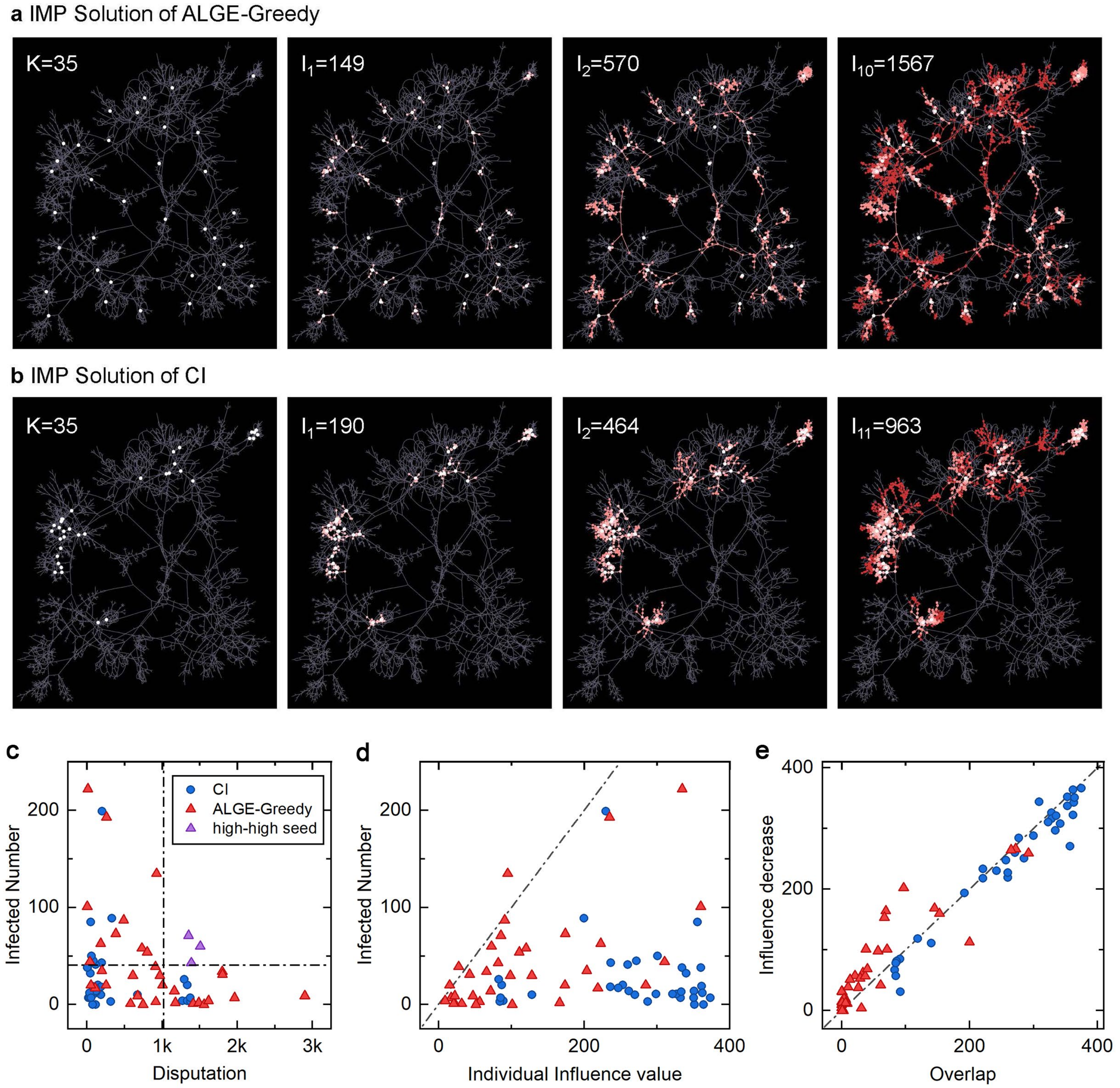

**Fig. 5 The propagation of ALGE-Greedy seeds in Influence Maximization Problem. a b** Propagation process of ALGE-Greedy and CI seeds in the Power network for IMP. The four charts respectively depict the scale of infection at time 0, 1, 2, and the final time. K represents the number of initial seed nodes (white nodes). $I_t$ represents the number of infected nodes at time t. **c** The disputation and final infected number of each seed from ALGE-Greedy and CI. Some seeds of ALGE-Greedy have both high disputation and high infected number over mean values (dot-and-dash lines) shown as high-high seed in purple. **d** The true individual influence value and actual infected number of each seed from ALGE-Greedy and CI. Many CI seeds with high individual influence have poor infected number in IMP simulation. **e** The relation between Influence decrease and overlap. Node Influence decrease means the decrease from the true individual influence value to the infected number in collective influence. Overlap value of nodes is calculated by BFS and the true individual influence values of all seeds.

# Discussion

In this paper, we have studied how to achieve precise influence evaluation across diverse networks. Although there have been numerous studies about network influence, there has been a lack of quantitative approaches to precisely measure the node influence considering network individualized information. Precise influence evaluation not only involves precisely ranking the nodes but also predicting the numerical values of their influence precisely. Such accurate and quantitative analysis method of characterizing networks is highly urgent and crucial for cutting-edge of network vulnerability[68] or control[69]. Traditional methods often rely on generic and qualitative algorithms that lack specificity and stability across different networks. These methods fail to precisely predict node influence due to lack of adjusting their computational approach based on the individualized structure of the network. In contrast, our method has the ability to enhance the precision by extracting individualized information of specific networks, based on machine learning method of active learning. Such a method adjusts the prediction model parameters incorporating the individualized structure of the network, enabling precise evaluation of node influence. With respect to fault, disease, and rumor propagation, this result allows us not only to proactively identify critical high-influence nodes as immunization targets before disaster occurring, but also to trace back to the source of infection based on the real impact of the event that have already occurred.

Our results indicate that, for a given structure, different algorithms could yield extreme inconsistency of influence evaluation. This is due to different perspectives of these algorithms, such as centrality, clustering coefficient, coreness, voting, local structure, weighted combinations of them, etc. These algorithms focus on heuristic calculation based on network structure metrics instead of evaluating influence itself. Therefore, some unusual structures hardly precisely described by generic metrics would render these methods ineffective high disputation. These unusual structures with individualized features pose critical challenge. While general indicators to some extent effectively characterize complex networks, those elusive structures without explicit discussion might be the final puzzle piece for thorough comprehension of system structure. This encompasses current hot researches such as absorption[70], hyperbolicity[71], ring structures[72], and may also involve some undiscovered "hidden structure", which just represent the identity features of each network. Only by understanding how individual nodes in a network organize their individualized features across diverse scales, can we perform various network classification task effectively with interpretability.

Current active learning methods on graph data generally learn the embedding representations of the unlabeled nodes by the model previously trained by labeled nodes to encode their potential information contribution to the model and then select the most informative candidates for training[73-75]. However, such learned representations might be invalid when the model is under-trained or there are only a few labeled nodes. Our active learning methods can directly select candidates based on the structural information of the unlabeled data without the need for model training and labeled nodes. Some methods also incorporate simple metrics such as graph centrality or influence as additional weights[76-78], which can only exploit topological information in a one-sided perspective. Indeed, nodes in graphs own extensive information from topology beyond centrality and influence.

Relying on the relative entropy measure[64] that combines both global and local information of nodes, we propose to construct node correlation networks as the foundation of an active learning method. This implies that differing from the original graph structure representing the interactions between nodes, we establish a high-dimensional mapping of graph data that reflects the high-order similarity among nodes, enabling us to balancing representativeness and informativeness intuitively by selecting both the most representative and hard samples for training[79], thus significantly improving the performance of graph learning task. Our proposed active learning paradigm for graph learning based on structure relative entropy of networks achieves topological clustering[80] of unlabeled data points through leveraging the structure information of network data. In detail, building upon the proposed topological correlation network, our approach can develop different structural similarity metrics and node identification algorithms tailored to specific active learning requirements for data selection. In downstream scenarios of graph learning, current active learning methods are designed for restricted applications, primarily suitable for classification problems and are not conducive to cross-network learning, whereas our approach has comprehensive applications to various scenarios and cross-network learning, such as regression, clustering, and more.

# Methodology

## Definitions

### Kendall's coefficient

Kendall's coefficient[45] is a measure to assess the consistency between ranking variables, which is commonly used to evaluate the accuracy of algorithms in predicting the ranking of node influences. For a sequence of nodes $V=(v_1,v_2,\cdots,v_N)$, we initially get their true influence rank sequence $X=(x_1,x_2,...x_N)$ where $x_i$ is the true influence rank of $v_i$ based on 1,000 independent simulations. Then we can get a prediction rank sequence $Y=(y_1,y_2,...y_N)$ where $y_i$ is the rank of $v_i$ based on the ranking algorithm. From $X$ and $Y$, a binary tuple $(X,Y)$ is obtained. Kendall's coefficient $\tau$ is calculated by the number of concordant and discordant pairs in $(X,Y)$:

$$\tau=\frac{2(n_+-n_-)}{N(N-1)}, \tag{1}$$

where $n_+$ and $n_-$ are the number of concordant and discordant pairs in $(X,Y)$ respectively and $N$ is the number of nodes.

### Disputation

The disputation of node $i$ is defined as the average absolute difference between its rank index of different algorithms and its true rank index, calculated by:

$$D_i = \frac{\sum_{j=1}^{m} | r_{ij} - r_{i0} |}{m}, \tag{2}$$

where $r_{ij}$ is the $j^{th}$ algorithm rank of node $i$, $r_{i0}$ is the true rank of node $i$, and $m$ is the number of available algorithms. And we choose 8 existing algorithms (k-shell[37], LID[42], NCVR[43], DCL[44], H-index[45], CI[49], RCNN[57], GLSTM[58]) to calculate the disputation. To ensure the generalization of the disputation for existing algorithms, we would exclude the largest prediction error among the 8 algorithms when calculating the disputation of each node.

**Quantified overlap**

We quantify overlap through the true individual influence of seeds. Initially, based on the true individual influence values of seeds, we employ the BFS algorithm to identify potential influence child-nodes of all the seeds. Subsequently, for one seed in the seed set, we calculate the union of child-nodes for the other seeds. The intersection size between the child-nodes of the current seed and the union of child-nodes for the other seeds represents the overlap value for that seed node. The overlap value of seed $i$ is calculated by:

$$O_i = \left| C_i \cap \left( \bigcup_{j \neq i}^{K} C_j \right) \right|, \tag{3}$$

where $C_i$ is the child-node set of seed $i$, $K$ is the seed set size.

In order to analyze the impact of overlap on collective influence, we calculate the influence decrease from true individual influence value to actual infected number in collective influence by:

$$ID_i = \max(0, I_i - IN_i) \tag{4}$$

where $ID_i$ is the influence decrease of seed $i$, $I_i$ is the individual influence of seed $i$ and $IN_i$ is the actual infected number of seed $i$. And the actual infected number of seed $i$ means the number of all the child nodes initially transmitted from seed $i$.

## Simulation on synthetic networks for pre-training

Before pre-training the prediction model, we need to obtain the real influence of nodes by simulation as the task information input to train the learning model. We conduct simulations on some synthetic networks, employing the standard SIR model as the diffusion model. In the SIR model, the node influence is defined as the number of final recovered nodes, when a particular node is the initial infected node and all other nodes are susceptible. In our simulation, the infected rate $\beta = 1.5\beta_c$, where the recovered rate $\lambda = 1$, $\beta_c$ is the epidemic threshold of SIR model. According to the heterogeneous mean-field theory[2,81], the approximate value of the epidemic threshold $\beta_c$ is given by:

$$\beta_c \approx \frac{\langle k \rangle}{\langle k^2 \rangle - \langle k \rangle}, \tag{5}$$

where $\langle k \rangle$ is the average degree of the network, $\langle k^2 \rangle$ is the mean square degree of the network. For each node, 1,000 independent runs are preformed, and its influence value is calculated by the average number of final recovered nodes.

## Influence prediction in real networks

In the pre-train part, we create a graph learning model to fit the graph structure features to the node influence in SIR model, and the graph learning model is first trained on some small synthetic networks aiming to have a preliminary ability to predict the node influence in diverse real networks. We construct the graph learning model based on the advanced graph attention network architecture GATv2[63] to predict influence. The graph learning model takes the network topology and node features as input and outputs the prediction values of the node influence. In our graph learning architecture, we utilize 3 convolutional layers with 8 heads for embedding, 2 fully connected layers for regression and Mean Squared Error (MSE) as loss function, to establish a mapping model from node features and network topology to node influence (refer to Supplementary Note 1 for details). The pre-trained graph learning model holds an initial mapping function between node features and their influence under the general structure features. Facing real networks, we employ an active learning approach based on graph entropy to fine-tune the prediction model parameters for transfer learning and improve the accuracy of predicting the node influence value.

After pre-training and active learning, our model integrates task information and structural information to achieve approximate prediction of node influence on the network. When the normalized features of one node are fed into the prediction model along with the network's topology, the estimated node influence values will be output.

## Active learning based on graph entropy

The active learning approach is an essential component for retraining the predictor to achieve precise quantification of node influence. During the AL process, we selected training nodes based on their relative entropy, which measures the similarity of a node to other nodes in terms of local and global structures. For detail, we first build the normalized correlation matrix $S$ based on the relative entropy[64,65]:

$$r_{ij} = RE_{ij} + RE_{ji}, \tag{6}$$

$$R = \begin{pmatrix} r_{11} & \cdots & r_{1n} \\ \vdots & \ddots & \vdots \\ r_{n1} & \cdots & r_{nn} \end{pmatrix}, \tag{7}$$

$$s_{ij} = 1 - \frac{r_{ij}}{\min_{i,j \in V}(r_{ij})}, \tag{8}$$

$$S = \begin{pmatrix} s_{11} & \cdots & s_{1n} \\ \vdots & \ddots & \vdots \\ s_{n1} & \cdots & s_{nn} \end{pmatrix}, \tag{9}$$

where $RE_{ij}$ is the relative entropy of node $i$ to node $j$, $r_{ij}$ represents the correlation between node $i$ and node $j$ based on relative entropy, $V = \{1, 2, \cdots, n\}$ is the node set, $R$ is the correlation matrix, and $S$ is the normalized correlation matrix. And larger $s_{ij}$ means nodes $i$ and $j$ are more similar.

Then, we establish a correlation network based on the correlation matrix. In the correlation network, the edges are determined by an edge threshold; if the correlation between nodes is higher than the threshold, an edge exists, otherwise, it doesn't. We employ the Bisection method to determine the network's edge threshold. The threshold must satisfy that the corresponding correlation network is connected and has the fewest edges. Finally, we iteratively selected representative nodes from the correlation network. After selecting representative nodes based on the structural information of the real network, their influence is calculated by simulation. Then, these labeled nodes are used to fine-tune the prediction model, enabling the better prediction of node influence values on the real network (refer to Supplementary Note 2 for details).

## Solve the influence maximization problem

The influence maximization problem can be approximated as a set cover problem or an integer programming problem with estimated influence values for all nodes in the network obtained from precise influence evaluation. For detail, we first select the highest influence node as the first seed and nodes in its approximate influence ranges are marked as all nodes influenced by this seed. That means all other nodes cannot infected these marked nodes and thus their current influence ranges should remove nodes in marked node set. Based on current influence ranges of residual nodes, we update their influence and select the highest influence node among them as the next seed. The selection and update processes alternate and the termination condition is that the resource limit or seed number limit has been reached (refer to Supplementary Note 3 for details).

## Data availability

The data of networks in this study is available at https://github.com/Zhu-BY/ALGE.

## Code availability

Computer code for the ALGE framework proposed are available on GitHub at https://github.com/Zhu-BY/ALGE.

## Acknowledgements

This work was supported by the National Natural Science Foundation of China (Grants 72225012, 72288101, 71822101 and 71890973/71890970), the Fundamental Research Funds for the Central Universities.

## Author contributions

B.Z. and D.L. designed the main idea of the research; B.Z., Q.S. and J.L. designed the ALGE framework; B.Z. performed the experimental evaluation. All authors contributed to discussing the results and writing the manuscripts.

## Competing interests

The authors declare no competing interests.

## Additional information

Correspondence and requests for materials should be addressed to D.L.